\documentclass{osa-article}

\journal{osajournal}
\articletype{Research Article}

\def\e{\begin{equation}}
\def\f{\end{equation}}
\def\-#1{{\bf #1}}
\def\o{\omega}

\def\l#1{\label{eq:#1}}
\def\r#1{(\ref{eq:#1})}

\begin{document}

\title{On subwavelength resolution granted by dielectric microparticles}

\author{ R.~Heydarian,\authormark{1,*}  and C.~Simovski\authormark{1,2}}
\address{
\authormark{1}Department of Electronics and Nano-Engineering, Aalto University, P.O. Box 15500, FI-00076 Aalto, Finland\\
\authormark{2}Faculty of Physics and Engineering, ITMO University, 199034, Birzhevaya line 16, Saint-Petersburg, Russia}
\email{\authormark{*}reza.heydarian@aalto.fi} 

\begin{abstract}
In this work we report a theoretical study of the lateral resolution granted by a simple glass microcylinder. 
In this 2D study, we had in mind the 3D analogue -- a microsphere whose ability to form a 
deeply subwavelength and strongly magnified image of submicron objects has been known since 2011. Conventionally, 
the microscope in which such the image is observed is tuned so that to see the areas behind the microsphere. This corresponds 
to the location of the virtual source formed by the microsphere at a distance longer than the distance of the real source to the miscroscope. 
{Recently, we theoretically found a new scenario of superresolution, when the virtual source is formed in the wave beam transmitted through the microsphere. 
However, in this work we concentrated on the case when the superresolution is achieved in the impractical imaging system, in which the microscope objective lens is replaced by a microlens located at a distance smaller than the Rayleigh range.} The present paper theoretically answers an important question: 
which scenario of far-field nanoimaging by a microsphere grants the finest spatial at very large distances. 
We found that the novel scenario (corresponding to higher refractive indices) promises further enhancement of the resolution. 
\end{abstract}

\section{Introduction}

Before 2011 the far-field subwavelength imaging without fluorescent labels (key component in the stimulated emission depletion method) and sophisticated post-processing was basically related to an effect of the so-called metamaterial hyperlens. This is a tapered/curved nanostructure with alternating plasmonic metal and dielectric constituents (both strongly submicron ones) forming the so-called hyperbolic metamaterial \cite{Hyper1,Hyper2,Hyper3,Hyper4}. The situation changed after the discovery 
of the same but drastically improved functionality offered by a simple glass microsphere on a silicon substrate \cite{Hong}. In most of studied structures the dielectric microspheres whose refractive index was within the interval $n=1.4-1.8$ demonstrated the broadband {spatial resolution} of two point scatterers separated by the subwavelength gap $\delta\ll \lambda$, accompanied by the magnification of this gap $M\gg 1$. Many such cases were studied already in the initial paper \cite{Hong}. In that work the best achieved result was $\delta_{\rm min}=\lambda/8$ and $M=50$. Further experimental studies of this effect revealed many interesting features (see in \cite{Lecler,Kassamakov,Astratov,Zhou,Maslov,Cang,NC}). The minimal spatial resolution $\delta$  of two point scatteres located either on the sphere surface (practically, sandwiched between the sphere and substrate) or in its near vicinity (practically, in a crevice between the bottom surface of the sphere and the substrate) noticeably depends on the refractive index of the microsphere and on the illumination. In the broadband regime this resolution is inverse proportional to the image magnification $M$. The results for ultimate values of $\delta$ and $M$ are different for coherent and non-coherent illumination, for polarized and non-polarized light, they also depend on the incidence angle if the imaging is obtained in the laser light. However, even in the daylight a simple glass microsphere with the refractive index in the interval $n=1.4-1.6$ demonstrated the broadband subwavelength resolution $\delta\ll \lambda$ and magnification $M\gg 1$.  

One theoretically found several resonant mechanisms of the far-field magnified subwavelength imaging functionality (that can be called for brevity, hyperlens functionality) of the dielectric microsphere. Some of them are related to the resonances of a spherical cavity, such as whispering gallery resonances and dipole Mie-resonances (see e.g. in \cite{Lecler,Zhou,Maslov}). One has shown experimentally and explained qualitatively that $\delta$ and $M$ can be improved utilizing a plasmon resonance if the sphere or the substrate are covered by a nanolayer of plasmonic metal \cite{Cang}. In these resonant cases one achieves $\delta=(0.07-0.1) \lambda$. 
One has proved that the effect $\delta\ll\lambda$ survives when one incorporates the microsphere into a polymer film so that to rapidly obtain the large-area imaging shifting the array of spheres like a solid body \cite{Astratov}. However, a satisfactory explanation of the broadband, non-resonant subwavelength imaging granted by a microsphere was absent for a long time. Work \cite{Yang} which pretended to explain it via evanescent-to-propagating waves conversion and the reciprocity theorem was severely disputed in \cite{Astratov1}. 

It seems strange why this lack of theory was possible for a so promising imaging device. Really, about a much worse and much more expensive metamaterial hyperlens
many dozens of theoretical papers were published. And only one known attempt \cite{Yang} was done before 2020 so that to explain how 
the non-resonant glass microsphere may create a subwavelength image. Perhaps, it is so because to simulate the whole imaging structure 
is strictly speaking impossible. One may simulate only the first stage, in which the incident light scattered by an object  
transmits through the sphere, forming the wave beam imaging this object. And even this simulations demands huge computational resources because the hyperlens 
functionality of the sphere corresponds to the case when $R\gg\lambda$. Moreover, to properly simulate even this first stage for an amount of different microspheres
huge computational resources should be spent. Therefore, in \cite{Astratov1} microspheres were replaced by microcylinders, and the dipoles by the dipole lines. This substitution offers a huge economy of the computational resources keeping the same underlying physics.

On the next stage of evolution the imaging beam created by an object behind the sphere experiences the Abbe diffraction (diffusion of power along the phase front). It holds over a giant optical path from the sphere to the microscope. Also, on the third stage the imaging beam passing the objective lens, becomes convergent and is focused.  Full-wave simulation of these processes is absolutely non-realistic, and to take the diffraction into account one utilizes the point-spread function technique, as the authors of \cite{Astratov1} have done. However, if the features of the subwavelength magnified imaging in the imaging beam are tiny there is a risk to lose these features in the approximate simulations. Therefore, one needs to find robust pronounced features in the imaging beam which would clearly point out the hyperlens functionality of the microparticle. 

In our recent works \cite{Reza,Reza1} the hyperlens operation of a dielectric microsphere beyond the resonances was related to the normal polarization of the scattering object. The simulated imaging beam produced by such a dipole already on the first stage of evolution manifests the prerequisites of the subwavelength imaging. These features lack in the simulations \cite{Astratov1} because the authors do not consider the normal polarization. We found an only work \cite{Wen} where the resolution 
granted by a microsphere for two normally polarized dipoles was compared to the case of tangential dipoles. In \cite{Wen} the superresolution $\delta=0.24\lambda$  was obtained for normally polarized dipoles in the full-wave simulation (stages 1 and 2) combined with the point-spread function technique.
This result was, however, achieved only for a microsphere almost touches a hemispherical microlens and they are coupled via near fields.
In this case evanescent waves are essentially involved in the imaging and even for the tangential dipoles one obtained $\delta=0.34\lambda$. 
Our claim is different: for normally polarized dipoles and only for them the superresolution granted by a glass microsphere is achievable without evanescent waves. In work \cite{conference} we explain that the normal polarization of the object occurs even in the case when the light is normally incident on the substrate (in the crevice 
between the sphere  and the substrate the cross-polarization results from the structural asymmetry).  

Paper \cite{Reza} refers to the coherent imaging in the obliquely incident laser light.
The non-coherent case is considered in \cite{Reza1}. In fact, already in \cite{Astratov} it was assumed that the microsphere excited by a point dipole located on its surface creates the virtual source (VS) located in front of the sphere. The substantial distance from the VS to the sphere center implies the magnification of the gap between two dipoles located on the sphere. Initially, we only aimed to prove that this idea is correct i.e. that the wave beam created by a point source after its transmission through the microsphere is a spherical wave whose virtual phase center is distanced from the source and can be treated as a VS. 

In line with \cite{Astratov1} we have checked this hypothesis for the 2D case and found that the assumption from \cite{Astratov} 
is incorrect for a tangentially polarized dipole. The imaging beam in this case is very non-homocentric (except, perhaps, resonant cases that we did not analyse). 
This result fits that of \cite{Astratov1}. However, for a normally polarized dipole we obtained two opportunities for the imaging beam to be approximately homocentric. 
Approximate homocentrism means that the beam phase center is though not point-wise but is spread in a much smaller domain than the distance from this center to the glass microparticle on which the dipole is located. Such a non-ideal homocentrism still allows the subwavelength resolution. 

The main result of \cite{Reza1} is prediction of the new scenario of nanoimaging. Besides of the scenario assumed in \cite{Astratov} when the VS is in front of the sphere we found that for large $n$ and large $R/\lambda$ the imaging beam is collimated behind the sphere.  
The explicit interval of $n$ for which the subwavelength resolution is achieved in this scenario depends on the size parameter $kR\equiv 2\pi R/\lambda$. 
This scenario was numerically implemented for 2D microparticles (microcylinders) with $R\ge 10\lambda$ and $n\ge 1.44$. The 
conditions of the subwavelength imaging $\delta<0.5\lambda$ were obtained for $R\ge 15\lambda$ if $n\ge 1.49$. The imaging beam of parallel rays propagates 
almost without divergence until the distances of the order of the Rayleigh range $D_R$ and then sharply diverges transforming into a spherical wave.
This wave has the directional intensity pattern with the zero on the beam axis (the beam axis is the direction of the source dipole). 
Its phase center treated as the VS is spread, however, its effective transverse size is much smaller than $D_R$, that results in the high magnification $M$ allowing to resolve small $\delta$.

Both numerically revealed scenarios -- the known one when the VS is formed in front of the sphere and the new one when it is located behind its back side --
enable the superresolution ($\delta<0.5\lambda$) without involvement of evanescent waves. In fact, there is nothing extraordinary in the linear label-free superresolution without evanescent waves. Besides of  metalenses utilizing spatial superoscillations \cite{Zheludev}, there exists also Gustafsson's or structured-illumination microscopy.
In this method the numerical aperture of the microscope objective attains $2$ due to the use of an objective diaphragm with pinholes (see e.g. in \cite{Imaging}). 
Then the diffraction limit modified from $\delta=0.5\lambda$ to $\delta=0.5 f\lambda$, where the f-number $f$ can be as small as 0.5 (and even smaller \cite{Review}).
However, narrow pinholes make the intensity in the image very low,  the superresolution is restricted by optical noises \cite{Narimanov}, and the necessity to suppress these noises strongly restricts the applicability of such microscopes. In our case, the superresolution is achieved without sacrificing the intensity -- just due to the annular pattern of the imaging beam. Really, an exact zero of the electromagnetic field on its axis precisely points out the point source location.  
 
The present paper is dedicated to the comparison of two scenarios of nanoimaging granted by a glass microsphere.  
In \cite{Reza1} we concentrated our efforts on a proof that the microsphere superresolution is achievable in a single full-wave simulations -- without splitting 
the numerical model onto separate problems and using the point-spread function. We simulated the system where the parallel imaging beam does not diverge but is focused by another microlens placed at the distance shorter than $D_R$. However, this nanoimaging system grants a modest magnification and is microscopic i.e. impractical. A practical system -- that with a macroscopic objective lens and a giant optical path of the imaging beam -- was considered only qualitatively. In the present paper, we do it more precisely. Our goal is not to obtain the best possible resolution. We aim namely to understand which of two scenarios of the microsphere nanoimaging is more promising for superresolution.

\section{Superresolution of real sources via resolution of virtual sources}
 
\subsection{Simplistic retrieval of virtual sources positions}
 
A point dipole located on the surface of the sufficiently large microsphere or in its close vicinity creates inside the sphere the field which can be identified with the continuum of rays propagating from the dipole and experiencing the partial or total internal reflection from its surface. The diffraction effects and excitation of creeping waves 
are not negligible, as such. However, the transmitted wave beam -- that behind the large sphere -- is basically formed in accordance to the ray optics \cite{Reza1}. 
If the refractive index is large enough i.e. $n>1.44$, the transmitted beam turns out to be collimated \cite{Reza1}.   
At the distance nearly equal to $D_R$ (in the 3D geometry $D_R\approx \pi R^2/\lambda$, and in the 2D case $D_R\approx 2 R^2/\lambda$) the 
collimated beam starts to diverge and at the distances larger than $D_R$ becomes nearly the spherical wave. The intensity pattern of this wave is very directional due to two factors. First, the wave results from the beam of finite-width and at large distance from the beam axis its intensity exponentially decays. Second, the dipole does not radiate along the beam axis $y$ i.e. the divergent beam keeps hollow. 

\begin{figure*}[htbp]
\centering
\includegraphics[width=14cm]{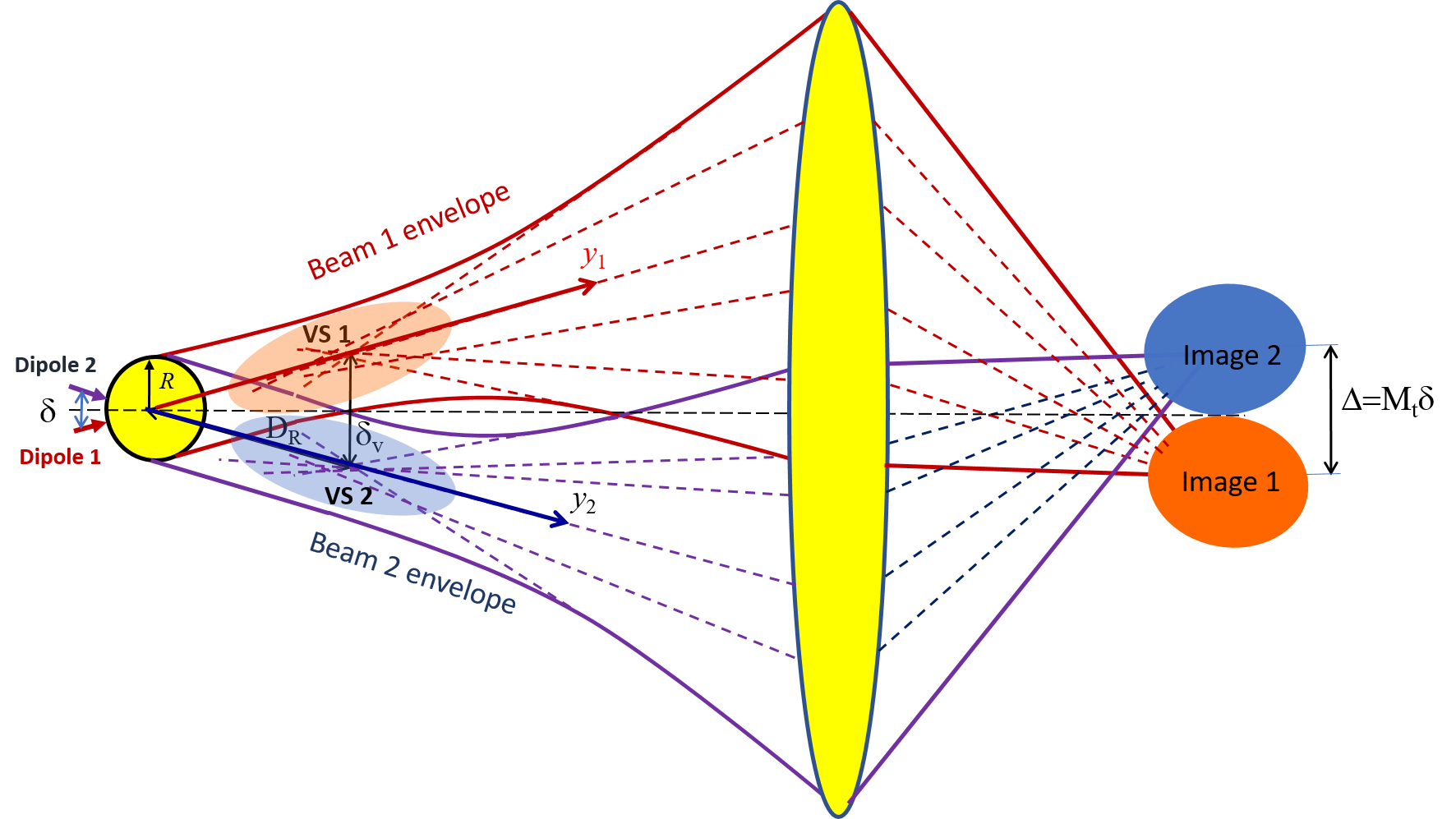}
\caption{Schematic illustrating to the simplistic ray-tracing concept of the virtual source (dashed lines are normal to the wave fronts of two diverging beams created by non-coherent point dipoles 1 and 2). If we adopt that the ultimate resolution corresponds to the case when the gap $\delta_V$ between the centers of two VSs is equal  $d_V$ the ultimate resolution equals to $\delta=\lambda/\pi$.} 
\label{Pic1}
\end{figure*}

If our divergent beam had exactly spherical wave front, the virtual source (VS) would have been point-wise and the spatial resolution granted by the microsphere would have been ideal. Of course, it is impossible and full-wave simulations of \cite{Reza1} have shown that the phase center of the divergent imaging beam is spread. It more spreads along the beam axis $y$  that implies poor spatial resolution in the longitudinal direction. However, across the beam the phase center spread is
smaller than the beam effective diameter ($2R$). In the simplistic model illustrated by Fig.\ref{Pic1} the dimensions of VS is found using the simplest variant of the ray tracing technique. Namely, taking a wave front of the diverging beam we plot the set of perpendicular lines to it, and the domain where they intersect determines the VS. Following to this procedure we could estimate in \cite{Reza1} the effective diameter of the VS $d_V\approx R$. 

Placing two point dipoles with the subwavelength gap $\delta$ at the left side of the sphere we obtain two VSs at the right side. In Fig.\ref{Pic1} we schematically show the case when there is no intersection between two VSs. However, they may partially overlap and still can be resolved. 
Assuming that the center of the VS is distanced by $D_R$ from the center of the sphere, the distance $\delta_V$ between the centers of two VSs 
can be found as $\delta_V=\delta D_R/R$. In order to estimate the spatial resolution in work \cite{Reza1}
we assumed that the distance $\delta_V= d_V/2$ is the minimal one when two VSs still can be resolved by the microscope. 
This approach corresponds to the Rayleigh criterion that allows the intersection of two intensity maxima on the level 70\%.  
It is evident, that this simplistic model delivers the minimal spatial resolution $\delta_{\rm min}=\delta_V R/D_R=\lambda/2\pi$. 

However, the applicability of the Rayleigh criterion for our VSs is disputable. This criterion refers to the case when the wave beam has the maximum on the axis, whereas our imaging beam is hollow. We introduce another criterion of the ultimate intersection of two VSs when they are still resolvable. This criterion is illustrated by Fig.\ref{Pic2}(b) and (c).
Let us demand that the null of the VS1 (created by dipole 1) is not compromised by the local maximum of the VS2 (created by dipole 2), and vice versa. 
In this case the minimal allowed distance $\delta_V$ between the centers of the VSs is larger than $d_V/2$. 
Below we will find this ultimate intersection. The magnification granted by the microsphere is defined as $M=\delta_V/\delta$, where $\delta_V$ is the gap between the centers of two VSs. The total magnification $M_t$ determining the image size $\Delta=M_t\delta$ is the product of $M$ by the magnification of the microscope $M_m$.


Another drawback of the geometric approach to the search of the VS is assumption that the distance from the sphere center to the center of the VS is equal $D_R$.
In fact, it may be smaller or larger, $D_R=\pi R^2/\lambda$ (for the 2D case $2 R^2/\lambda$) is an accurate estimate only for a Gaussian beam, and for our hollow beam
it only gives a rough estimation. If we adopt our new criterion of resolution we have to rewrite the formula $\delta={R\delta_V/ D_R}$ in the form: 
\begin{equation}
\delta = {R\delta_V/ D}
\l{eq1}
\end{equation}
\noindent 
where $\delta_V\ne R$ and $D\ne D_R$ are parameters to be found as precisely as possible.

\subsection{Retrieval of virtual sources via backward propagation}
 
\begin{figure*}[htbp]
\centering
\includegraphics[width=13cm]{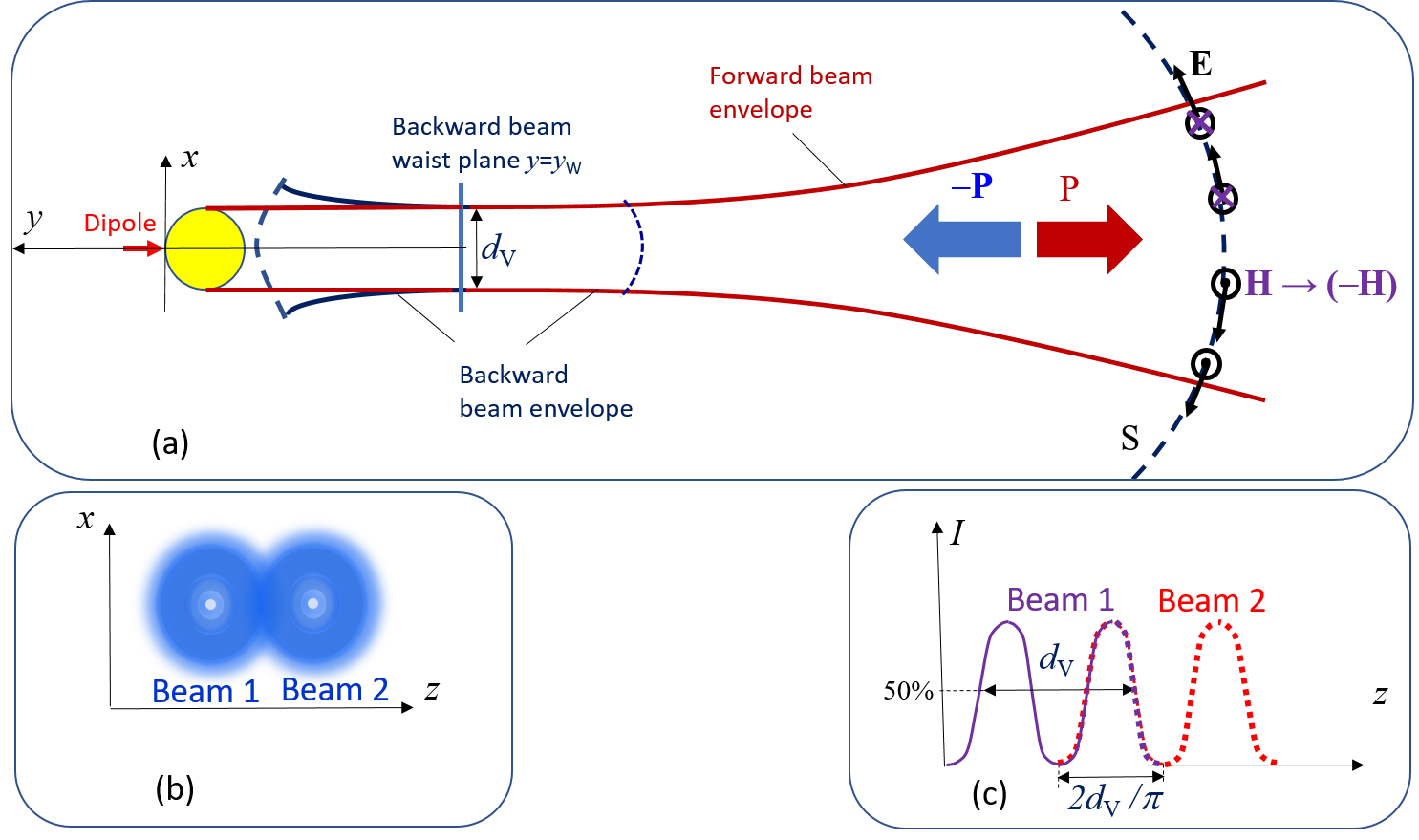}
\caption{(a) Sketch of the imaging beam and its backward implication. Dashed lines depict the wave fronts of the imaging and backward beams. Inversion of the magnetic field in the imaging beam at the surface $S$ inverts the Poynting vector $\-P$ and results in the backward beam. Electromagnetic field of this beam is transverse and in-phase in the plane of the waist. The disk of diameter $d_V$ can be identified with the VS. (b) Two partially overlapping VSs are reliably distinguishable if the field nulls on their axes are not compromised by the tails of the other VS maximum. (c) The intensity distribution across two partially overlapping VSs in this ultimate case.  
}
\label{Pic2}
\end{figure*}

A new concept of the VS is based on the idea of the equivalent free-space past of the imaging beam. Let us take a wave front $S$ of the imaging beam in the region 
sufficiently distant from the sphere, i.e. where the Abbe diffraction is weak and, therefore, both electric $\-E$ and magnetic $\-H$ fields are strictly tangential to $S$. We may use the distribution of the electromagnetic field on $S$ obtained in full-wave simulations as the equivalent distributed source since the Green's formula is the strict formulation of Huygens' principle. This way we may calculate the further propagation of the imaging beam -- the dipole and the sphere are not needed for it. 
Moreover, we may reconstruct the virtual past of this beam in which there is no dipole and sphere. For it may invert e.g. magnetic field ($\-H\rightarrow -\-H$) 
keeping the same $\-E$. This way we invert the Poynting vector $\-P\rightarrow -\-P$ and obtain the backward version of the imaging beam. This backward beam 
converges in the same region where the imaging beam diverges. As it is schematically depicted in Fig.\ref{Pic2}(a) the backward beam 
after its convergence is collimated and in the waist plane $y=y_W$ its phase front becomes ideally flat. 
Of course, after this waist the backward beam diverges. The electromagnetic field in the plane $y=y_W$ is effectively concentrated inside the disk 
with the diameter $d_V$. This disk can be identified with the VS.

Really, in accordance to Green's theorem if we do the same inversion $\-H\rightarrow -\-H$ in the waist plane of the backward beam, we may consider this field as an equivalent source of surface electric ($\-H$) and magnetic ($\-E$) polarization distributed in the waist plane $y_W$ . 
{{This source exactly reproduces the field of the imaging beam at the surface $S$ and everywhere behind. In other words, it reproduces the field developed by the objective lens. It is so, in spite of the fact, that the backward beam calculated in this way  
does not reproduce the true field distribution in the vicinity of the sphere. 
However, this difference is not an error of our method, it is, in fact, our governing idea.

The objective lens has a finite aperture. It does not see the lateral radiation of the dipole we aim to image. Our finite surface $S$ 
aims to capture that part of the radiation which is responsible for the image. The sidelobes of the imaging beam which do not contribute into the imaging process
are not crossed by our $S$. Of course, if we modify $S$ so that to apply the Huygens principle in its strict formulation i.e. make $S$ infinite 
our backward beam will reproduce the true field created by the dipole everywhere. However, we do not need this true field, because it does not deliver the proper size of the VS! Only the waist of the backward beam is fully self-consistent with the concept of the virtual source as a finite effective source located in free space! 
The magnetic and electric fields in the waist are equivalent electric and magnetic polarizations orthogonal to the optical axis, which produce in the objective of a microscope the same electromagnetic field as that developed by this objective in the original problem. So, returning back from the point $y=y_S$ (where $y_S$ 
is the point where $S$ crosses the optical axis) we in fact determine the VS whose effective lateral size is that seen by the microscope. 
}}

Now, let us see why the VS is namely the waist and not any other cross section of the backward beam.
This is so because only at the plane $y=y_W$ the effective diameter $d_V$ of the VS makes sense. 
In this plane and only in it the electromagnetic field of the backward beam is transverse. 
In this plane and only in it the wave front coincides with the cross section plane $xz$. Therefore, only for this plane we may properly 
define the lateral size $d_V$ through the intensity distribution.    

As to the resolution of two parallel partially overlapping imaging beams, we adopt the concept that two VSs are distinguished if their cross sections partially overlap as it is shown in Fig.\ref{Pic2}(b) and (c). It is the ultimate overlapping when the null of one beam is not yet spoiled by the maximum of another one. 
In Fig.\ref{Pic2}(c) the intensity distribution is depicted as a function of a transverse coordinate. We see that in the ultimate case of the resolution 
the right maximum of the left VS coincides with left maximum of the right VS. Then two nulls of two adjacent VSs are not compromised.
Assuming that the shape of the concentric maximum of our hollow imaging beam is Gaussian, it is easy to show that the distance between the nulls of two adjacent VSs 
in the ultimate case is equal $2d_V/\pi$. Here $d_V$ is the  
effective outer width of the waist shown in  Fig.\ref{Pic2}(b) that is taken on the level of 50\% intensity. 
Thus, the ultimate distance $\delta_V$ between two resolvable VSs is not $d_V/2$ as it was assumed in \cite{Reza1}, but 
nearly $1.27$ times larger. 

In fact, this criterion of the ultimate resolution gives not a true resolution of two point sources seen in a microscope. 
It is a pessimistic estimate, because it implies parallel axes of two imaging beams created by two adjacent sources. 
However, two imaging beams created by two point sources with the gap $\delta$ located on the sphere and normally oriented to it   
have different axes $y_1$ and $y_2$ as it is shown in Fig.\ref{Pic1}, and the angle between these axes is equal $\alpha=\delta/R$ . 
However, in this paper we do not aim to accurately calculate the spatial resolution. We aim to understand which scenario of superresolution
grants finer $\delta_{\rm min}$. For it we cannot use the Rayleigh criterion because it is not applicable to hollow light beams, and have 
to elaborate a new criterion, which would be physically reasonable, reliable and unique for both scenarios. 

Definitely, our idea of the VS treated as the waist of the backward beam keeps valid 
also for the imaging beam having the phase center in front of the sphere. 
In fact, we do not know in advance where the backward beam should have the waist -- behind the spatial region occupied by the sphere in the simulations of the imaging beam, in front of this region, or even inside it. However, if the coordinate of the waist $y_W$ is close to that of the point from which the imaging beam 
seemingly diverges, it will tell us that the geometrical optics is applicable with high accuracy to the imaging beam. 
Thus, the purpose of this paper is to compare the ultimate resolutions $\delta_{\rm min}$ calculated in the same way for two scenarios of superresolution: the '
conventional one and the novel one.

\subsection{Non-resonant microparticle: comparison of two scenarios}

In our simulations we replace the 3D sphere by the 2D (cylinder) and utilize the COMSOL Multiphysics as we did in work \cite{Reza1}. 
It grants a huge economy of computational resources and allows us to follow the evolution of the imaging beam up to $1000\lambda$.  
However, even this path may be not sufficient (so that $S$ ensures the robust result for $d_V$) if $R$ is very large. Really, the Rayleigh distance is proportional to $R^2$, and if $R=20\lambda$, $D_R=800\lambda$. However, we will see below that in order to properly find the VS we need to know the field at the distances much larger than $D_R$. Practically, it means that we have to study the cases when $R<20\lambda$. In all our simulations the wavelength is fixed ($\lambda=550$ nm) and we vary $n$ and $R$. 

In \cite{Reza1} it was found that for $R/\lambda<10-15$ the approximation of geometrical optics as a model of the imaging beam formation is not applicable. 
{Since the idea of the collimated imaging beam resulted in \cite{Reza} namely from the ray optics, we concentrated in   
\cite{Reza1} on the case $R>15\lambda$.} This choice did not allow us to simulate the beam evolution at the distances much larger than $D_R$. 
The present study demands such simulations. Therefore, we searched the needed imaging regime for microparticles with $(R/\lambda)<11$. 

Our non-resonant imaging mechanism demands a quasi-continuum for the internal pattern. It corresponds to many TM-modes around the particle perimeter with nearly equal amplitudes. However, varying the size parameter of our microcylinder in the range $4<(R/\lambda)<11$ we saw a lot of multipole Mie resonances manifesting in the standing wave patterns inside the microparticle. At these resonances the imaging beam cannot be formed because the phase front of the transmitted beam keeps features of the internal pattern and is far from being flat. Then in the transmitted beam we observe strong interference effects which do not disappear even at the distances much larger compared to $D_R$. This implies several parasitic images i.e. the microparticle becomes not beneficial but harmful for imaging \cite{Reza1}. 
In accordance to \cite{Maslov} Mie resonances can be beneficial for superresolution, but it refers only to low-order resonances which imply $(R/\lambda)<1$.
Studying the range $4<(R/\lambda)<11$ we should avoid Mie resonances.

When the internal field pattern looks as a quasi-continuum, i.e. the excitation of the microparticle is not resonant, the formation of the imaging beam occurs basically in accordance to geometrical optics \cite{Reza1}. Depending on the radius $R$ and refractive index $n$ there can be two regimes -- when the light beam created by a point dipole being transmitted through the microparticle is diverging and its  
virtual phase center $y=y_{GO}$ is located in front of the microparticle (a conventional scenario of superresolution \cite{Astratov}) and when the transmitted beam is non-divergent and keeps collimated until the point $y=y_{GO}<0,\, |y_{GO}|\gg R$ that is the virtual phase center of the imaging beam divergent part (our scenario illustrated by Fig. \ref{Pic2}). The transition from one scenario to another one occurs for given $R$ when we vary $n$. Therefore, we needed to find such size parameter $R/\lambda$ for which the resonances do not arise for any $n$ in the broad interval of values which will allow us to observe in our simulations the transition from the first scenario to the second one and to compare the corresponding ultimate resolutions calculated with Eq. \r{eq1} where $\delta_V=2d_V/\pi$ is the gap between two ultimately resolved VSs and $D=|y_W-R|$ is the distance from the VS (waist of the backward beam) to the center of the microparticle. 

\begin{figure*}[htbp]
\centering
\includegraphics[width=15cm]{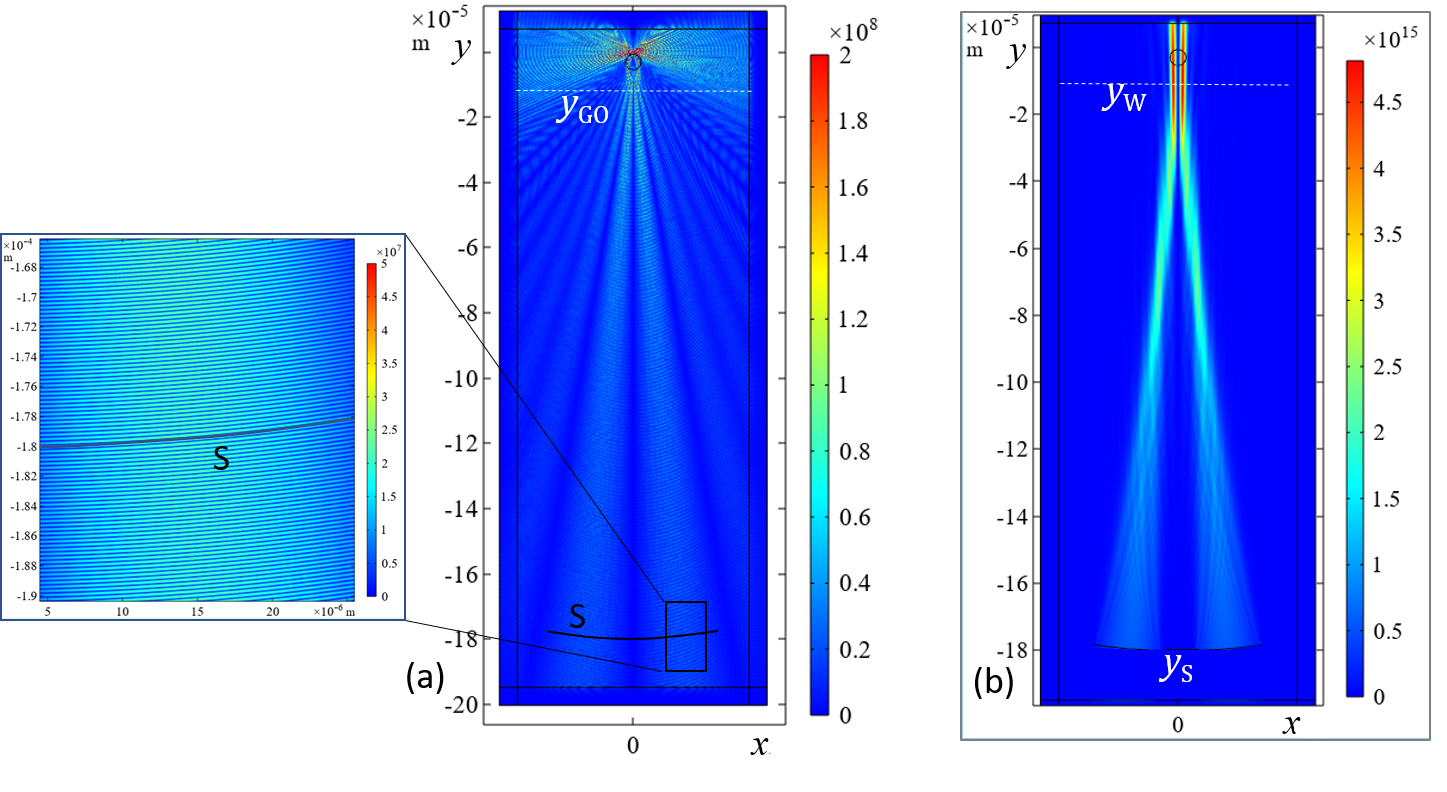}
\caption{(a) Instantaneous wave picture of the imaging beam for the case $R=4.3\lambda$, $n=1.46$. It is collimated until the plane $y=y_{GO}$ where the phase center of the diverging part is located (white dashed line). (b) Electric intensity color map of the backward beam simulated for free space, where the microparticle is shown 
(by black solid line) only for reference. In the waist plane $y=y_W\approx y_{GO}$ the phase front of the backward beam is flat and transverse.}
\label{Pic3}
\end{figure*}

We found that in the case $R=4.3\lambda$ the non-resonant regime holds if we vary $n$ from $1.35$ to $1.6$ with the step $0.01$. In other words, for 
$R=4.3\lambda,\, n=1.35, \, 1.36,\dots 1.6$ all the TM-modes excited by our dipole have the magnitudes of the same order and their interference mimics the internal quasi-continuum.  
As an illustration of our method we show the results for $n=1.46$. 
The electric field (amplitude and phase) color map is presented in Fig.\ref{Pic3}(a). The transmitted beam looks collimated until the plane $y_{GO}\approx -11\, \mu$m and diverges after this point rather sharply. Two main lobes of the diverging part have the phase center at $y=y_{GO}$. This distance is noticeably smaller than the Rayleigh range $D_R\approx 37\, \mu$m. This is not surprising since the Rayleigh theory was developed not for hollow beams.  

Starting from the distances $y=150-160\, \mu$m the wave front of the imaging beam becomes circular with very high accuracy. The inset of
Fig.\ref{Pic3}(a) 
presents the wave picture of the diverging beam in the enlarged scale. The surface (in our 2D case -- line) $S$ is shown in this color map and is an arc of a circle of radius nearly equal to $y_S-y_{GO}$.
Using the field distribution over $S$ we restore the backward beam whose intensity distribution is depicted as a color map in Fig.\ref{Pic3}(b).

The waist of the backward beam in Fig.\ref{Pic3}(a) turns out to be located practically in the same plane as the phase center of the imaging beam diverging part $y=y_W\approx y_{GO}$. The geometrical optics really works for 
the diverging part of the imaging beam, and the diffraction effects are minor! 
The distribution of the intensity in the waist plane corresponds to $d_V\approx 2.8\, \mu$m$=5.1\lambda$. The ultimate resolution turns out to be $\delta=\approx 1.46\lambda$. For $R=4.3\lambda$, $n=1.41\le n\le 1.54$ we have not obtained the resolution finer than $0.5\lambda$. However, it does not mean that the subwavelength resolution is not achievable for this interval of $n$ because our criterion of resolution is excessive. It only means that the intervals  
$n=1.41-1.6$ and $n=1.35-1.4$ grant better resolution. 

\begin{figure*}[htbp]
\centering
\includegraphics[width=15cm]{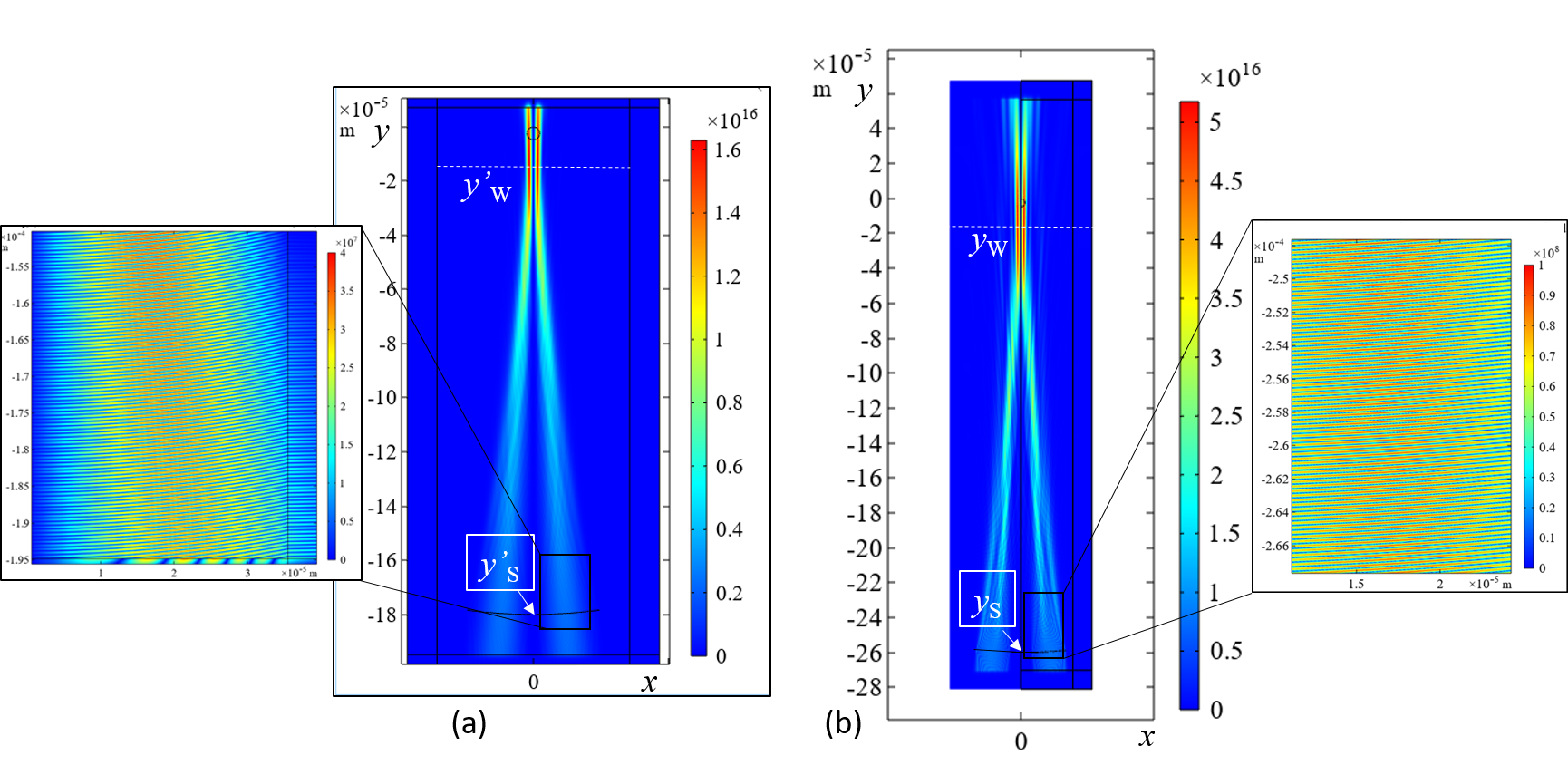}
\caption{Color map of the electric intensity for the case $R=4.3\lambda$, $n=1.6$. Backward beam above $S$ ($y>y_S$), imaging beam below $S$ ($y<y_S$).   
(a)  Wrong choice of $y_S=y'_S$ delivers wrong waist position $y'_W$. (b)  Correct choice $y_S= -260\, \mu$m delivers the correct waist position $y_W$. 
On the insets the instantaneous wave pictures of the backward beams are shown.  
}
\label{Pic4}
\end{figure*}

In order to properly find the waist coordinate $y_W$ one has to simulate the imaging beam until very large distances, 
larger than $D_R$ by an order of magnitude.  
If we take insufficient $|y_S|$ the waist location turns out to be dependent on $y_S$ i.e. the backward beam in the region of the waist 
is calculated wrongly. 
Fig.\ref{Pic4} illustrates this fact.  
Here we present the intensity color maps of the backward beam calculated in the case $R=4.3\lambda$, $n=1.6$ for two 
choices of $y_S$. If we take $y_S=y'_S= - 180\, \mu$m, as in Fig.\ref{Pic4}(a), the phase front in the enlarged inset looks pretty circular.
We checked that the vectors $\-E$ looks pretty tangential to such $S$  (vectors $\-H=\-z_0H_z$ are automatically tangential in our 2D problem). 
It must be so at so big distances (in this case $|y'_S|\sim 4D_R$). 
However, the distance $|y|=|y'_S|$ is still insufficient so that properly calculate the backward beam. 
Most probably, this is so due to quite high gradient of the field amplitude along the phase front as we can see in the inset. Taking 
$|y_S|= 180\, \mu$m we obtain the wrong waist position $y_W=y'_W\approx - 14\, \mu$m, that corresponds to the 
wrong ultimate resolution $\delta'_{\rm min}=0.75\lambda$. Meanwhile, taking the phase front at $y_S= - 260\, \mu$m
as it is depicted in Fig.\ref{Pic4}(b) we obtain $y_W\approx - 17\, \mu$m and the ultimate resolution is $\delta_{\rm min}=0.38\lambda$.
Further increase of $|y_S|$ does not change this result. Note that the gradient of the field amplitude along the phase front as it is seen in the 
inset of Fig.\ref{Pic4}(b) is not as high as in Fig.\ref{Pic4}(a). In our further studies we respect this requirement and choose   
$|y_S|$ sufficiently large. 

Note, that in Fig.\ref{Pic4}(a) and (b) the part of the imaging beam behind $S$ is shown in order to demonstrate that the backward beam 
in its converging part precisely reproduces the diverging part of the imaging beam. 

Varying $n$ in the limits $n=1.55-1.6$ we obtained $\delta_{\rm min}\le 0.5\lambda$ and $\delta_{\rm min}$ decreases monotonously versus $n$. 
The best result $\delta=0.38lambda$ corresponds to $n=1.6$.  If $n=1.45-1.54$ $\delta_{\rm min}> 0.5\lambda$ but the resolution also improves when 
$n$ increases. 
In all these cases $y_{GO}\approx y_W$.

\begin{figure*}[htbp]
\centering
\includegraphics[width=14cm]{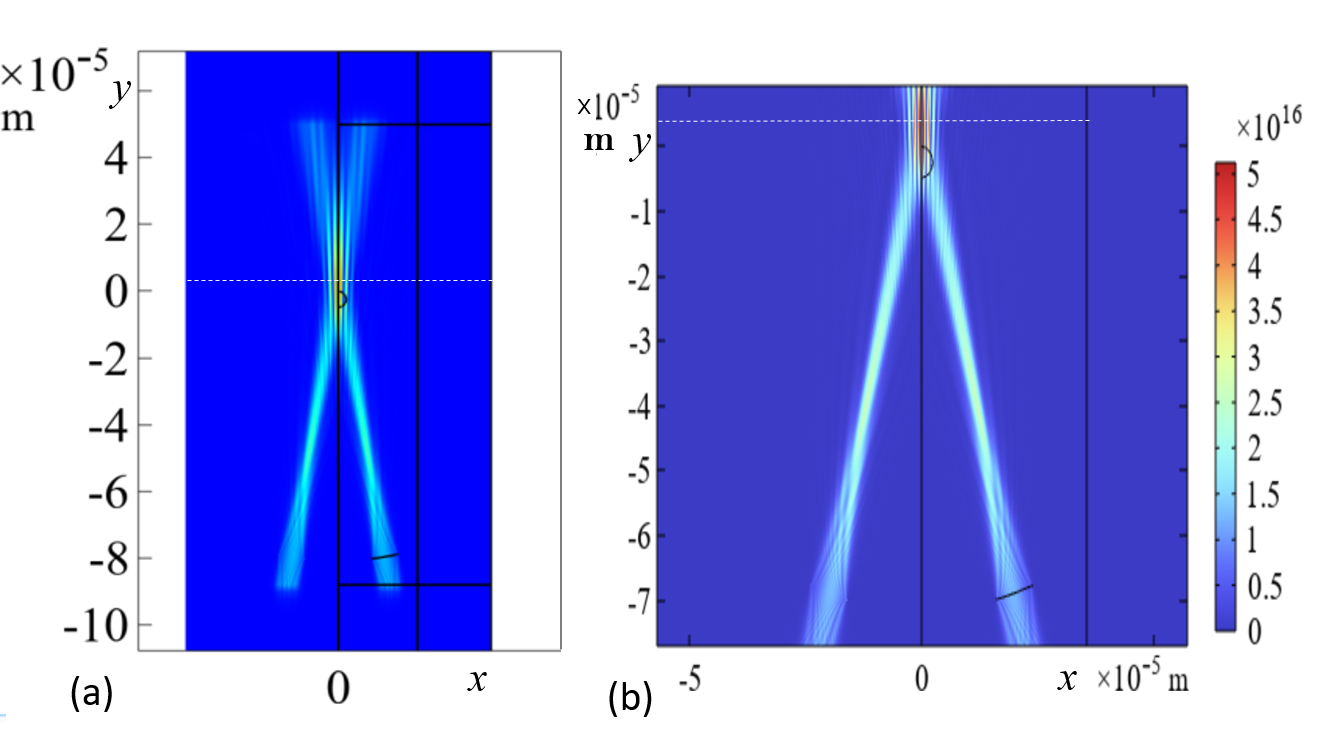}
\caption{Color map of the electric intensity for the backward beams in the cases $n=1.4$ (a) and $n=1.35$ (b). 
$R=4.3\lambda$ in both cases. Planes $y_W\approx y_{GO}$ are shown by dashed white lines. Surfaces $S$ are shown by black arcs.}
\label{Pic5}
\end{figure*}

If we take $n<1.44$ the known scenario of imaging predicted in \cite{Astratov} is implemented -- the VS is formed in front of the microparticle (at $y>0$). 
In Fig. \ref{Pic5}(a) we present the intensity color map of the backward beam (and of the forward beam in the region $|y|<|y_S|$) for $n=1.4$. Then $y_W=2.7\, \mu$m, 
$d_V=1.64\, \mu$m and $\delta_{\rm min}=0.89\lambda$. Further decrease of $n$ results in larger $y_W$ i.e. more distant location of the VS from the microparticle. 
It implies larger 
magnification of the gap $\delta$ and allows better resolution as well as the increase of $n$ if $n>1.45$ and the novel scenario is implemented. 
Fig. \ref{Pic5}(b) corresponds to the case $n=1.35$, when $y_W=3.0\, \mu$m, $d_V=1.01\, \mu$m and 
$\delta_{\rm min}=0.5\lambda$. We made simulations also for lower $n$ (though these values do not correspond to any practical solid material) 
and saw that the resolution slightly improves when $n$ decreases. For $n=1.3$ (water) we have obtained $\delta=0.48\lambda$.  

It allows us to assert that at least for the special case $R=4.3\lambda$ the imaging scenario revealed in our work \cite{Reza1} corresponds to finer superresolution compared to the conventional scenario. Our scenario also grants larger magnification of the VS, $M=\delta_V/\delta=2d_V/\pi\delta$ and, therefore larger total magnification $M_t=M M_m$. 
Though in this study we have obtained a quite modest superresolution, our values of $\delta$ are only pessimistic estimates (we do not know yet how to 
determine the resolution of two hollow beams taking into account the nonzero angle between their axes).

\subsection{Enlarging the microparticle}

\begin{figure*}[htbp]
\centering
\includegraphics[width=14cm]{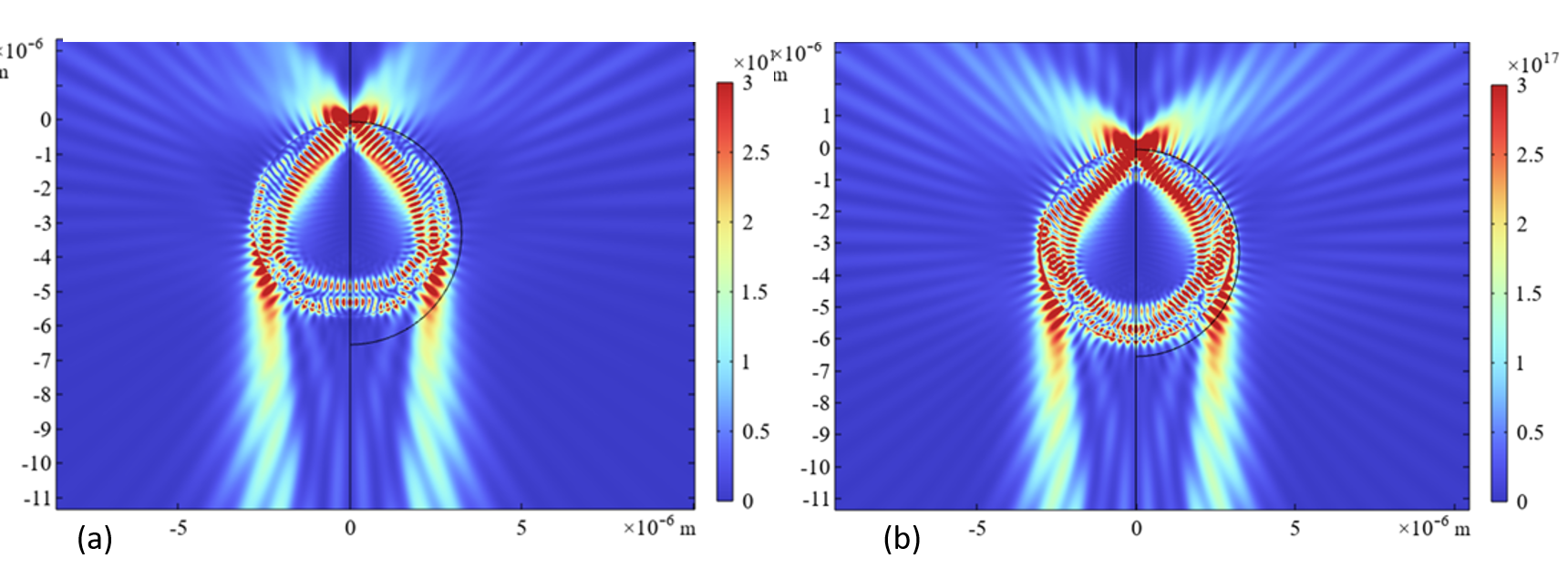}
\caption{Electric intensity distribution inside a microcylinder and in its vicinity for $n=1.6$, (a) $R=5.15\lambda$ and (b) 
$R=5.45\lambda$. The backward beam in both cases has no unique waist.}
\label{Pic7}
\end{figure*}

Aiming to improve the resolution we studied the cases when the radius of the microcylinder is enlarged up to $R=11\lambda$. Unfortunately, in this range of $R$ 
there is no possibility to vary $n$ for fixed $R$ keeping the quasi-continuum of the internal field (i.e. non-resonant regime). 
For $R=5\lambda$ the internal distribution seems non-resonant for $n=1.3-1.35$ and $n=1.55-1.6$. For the first interval of $n$ we observe the conventional regime with the VS in front of the microparticle, and for the second interval the transmitted imaging beam looks collimated until the distance of the order of $D_R$. In this meaning, the case $R=5\lambda$ does not differ from the previous one. For larger $R$ the situation is similar -- for sufficiently small 
$n$ the conventional scenario of imaging is implemented and for sufficiently large $n$ -- the novel scenario. Between these intervals there is a range of $n$ where 
the VS is located in the region of the microparticle, and these values of $n$ cannot allow the superresolution. 
There are also specific combinations of $R/\lambda$ and $n$ at which we observe the resonant pattern inside the microparticle, and in this case the imaging beam has numerous lobes just behind the microparticle. If we fix $n=1.6$ in the range $5\lambda<R<6\lambda$ this situation holds e.g. for $R=5.15\lambda$ and 
$R=5.45\lambda$. The intensity distributions for these two cases in the region of the microparticle is shown in 
Figs. \ref{Pic7}(a) and (b), respectively. Here we may distinct few resonant modes of close orders (one of them is whispering gallery mode). 
We see the strong interference pattern in the transmitted beam that results in several lobes of nearly same intensity even at the distances larger than the Rayleigh range.
The backward beam in these cases also splits onto several lobes and has several waists i.e. the imaging is not possible. 
In fact, the similar multiresonant regime exists for specific values of $n$ even in the case $R=4.3\lambda$ but we did not report above these exceptional cases.

\begin{figure*}[htbp]
\centering
\includegraphics[width=15cm]{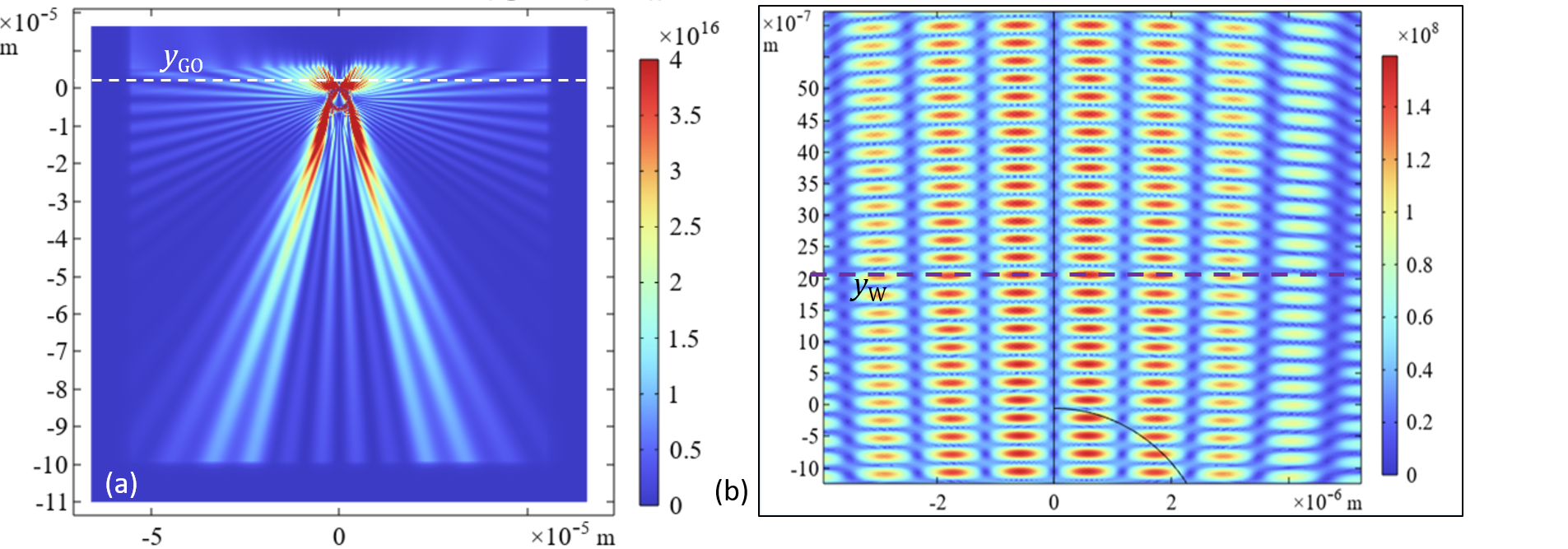}
\caption{(a) Intensity color map for the case $R=5\lambda$, $n=1.3$.  
(b) Instantaneous electric field color map for the backward beam in the domain of the waist. Planes $y_{GO}$ and $y_{W}$ are shown by dashed lines.
Black solid line in (b) shows the region of the microparticle.}
\label{Pic2new}
\end{figure*}

However, for $5\lambda<R<6\lambda$ the resonant regime holds for many values of $n$. Though for $R=5\lambda$ $n=1.3-1.4$ and $n=1.6-1.7$ the internal pattern looks 
like a quasi-continuum, the imaging beam splits onto pronounced lobes. It  
 clearly points out the leaky modes interference. 
Thus, the quasi-continuum inside the microparticle is only seeming. 
The case $R=5\lambda$ $n=1.3$ is presented in Fig. \ref{Pic2new}(a). 
We see several lobes in the imaging beam and it is not surprising that the backward beam has the same amount of lobes. 
However, its waist is unique in spite of the lobes which has the common coordinate $y_W$ 
where their phase front is almost flat. Moreover, this plane 
is predicted by geometrical optics with surprising accuracy.
The electric field color map is shown in Fig. \ref{Pic2new}(b). The waist part is located in front of the microparticle. 
The analysis of the phase front shows that the front is maximally flat at $y_{W}=+2.1\, \mu$m$\approx y_{GO}=2.2\, \mu$m.
The effective waist width is $d_V=2.7\, \mu$m and the ultimate resolution in our definition is equal
$\delta_{\rm min}\approx 1.9\lambda$. 

\begin{figure*}[htbp]
\centering
\includegraphics[width=14cm]{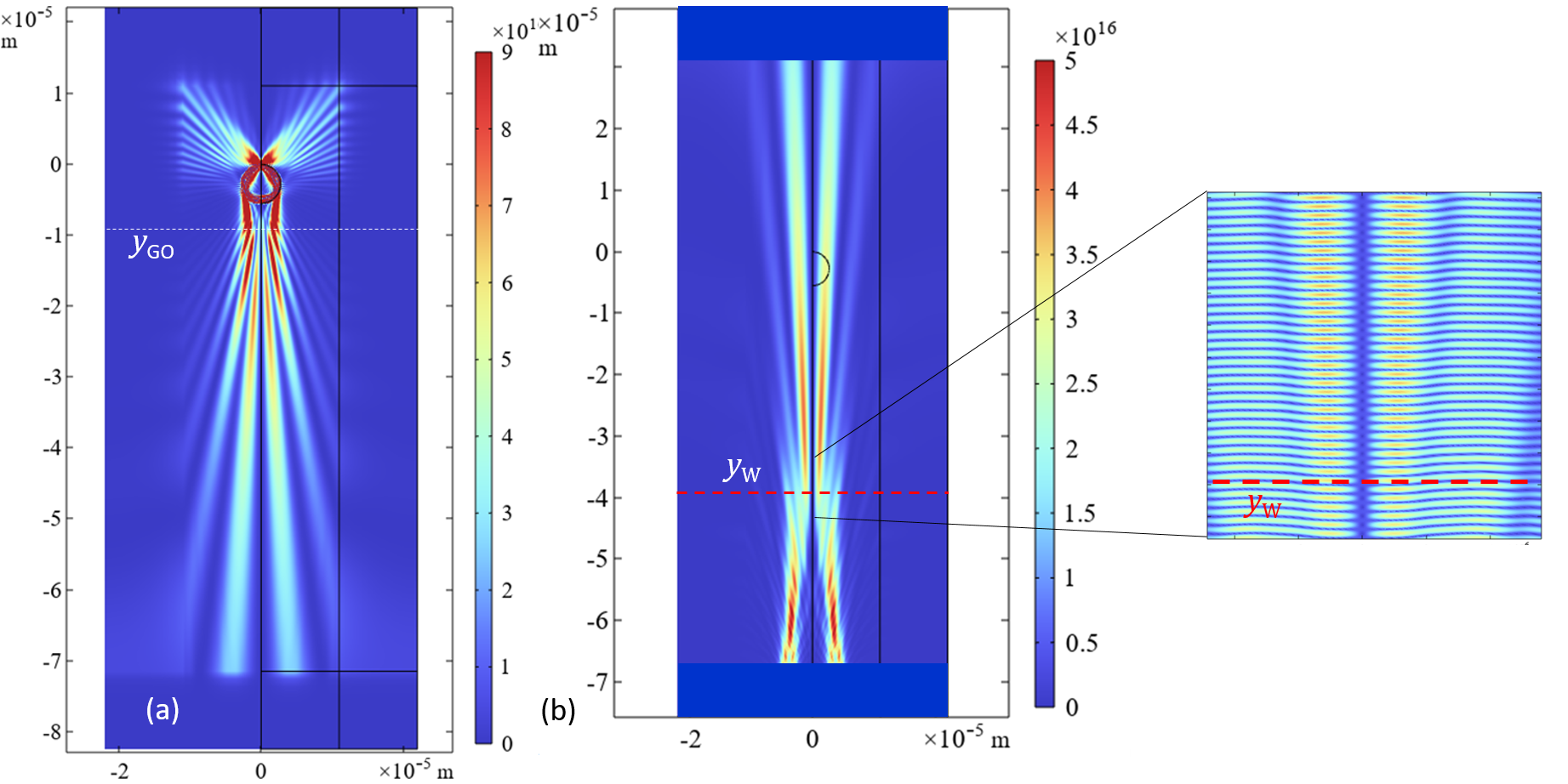}
\caption{(a) Intensity color map for the case $R=5\lambda$, $n=1.6$.  
(b) Instantaneous electric field color map for the backward beam in the domain of the waist.}
\label{Pic1new}
\end{figure*}

The case $R=5\lambda$, $n=1.6$ in Fig. \ref{Pic1new}(a) also looks non-resonant if we inspect the internal field pattern. 
The imaging beam is collimated qualitatively in accordance to the geometrical optics and diverges having the phase center at 
$y_{GO}\approx -10\, \mu$m. However, in its diverging part we again observe non-negligible lateral lobes. However, in spite of these lobes the backward beam 
has no lobes in the region of the waist.
The waist plane $y=y_W=-39.5\, \mu$m is uniquely determined via the requirement of the flatness of the phase front and orthogonality of the electromagnetic field in this plane. Note that the waist does not coincide with the plane of the maximal intensity concentration ($y=-60\, \mu$m). In the last plane the backward beam is still converging. The waist also does not coincide with the phase center of the imaging beam diverging part, that also implies the impact of the modal structure of the internal field though it is not detected visually.  In spite of these parasitic effects we obtain the ultimate resolution $\delta_{\rm min}=0.49\lambda$ that is much better than that in the case when $n=1.3$. 
This is so because the VS is much more distant from the real source compared to the conventional regime and the magnification $M$ is much larger.

\begin{figure*}[htbp]
\centering
\includegraphics[width=13cm]{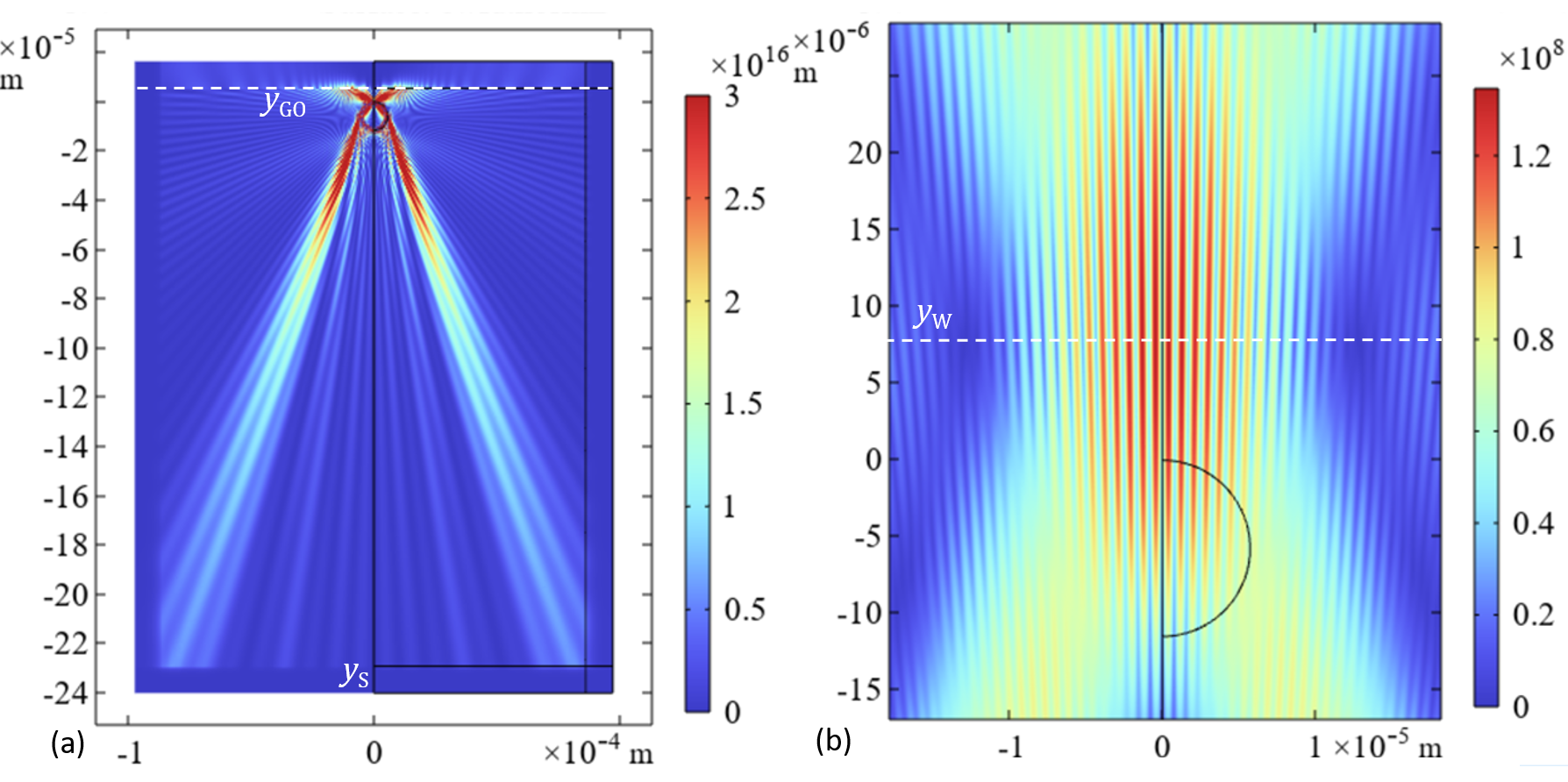}
\caption{(a) Intensity color map for the case $R=10.45\lambda$, $n=1.3$.  
(b) Instantaneous electric field color map for the backward beam in the domain of the waist.}
\label{Pic4new}
\end{figure*}

\begin{figure*}[htbp]
\centering
\includegraphics[width=13cm]{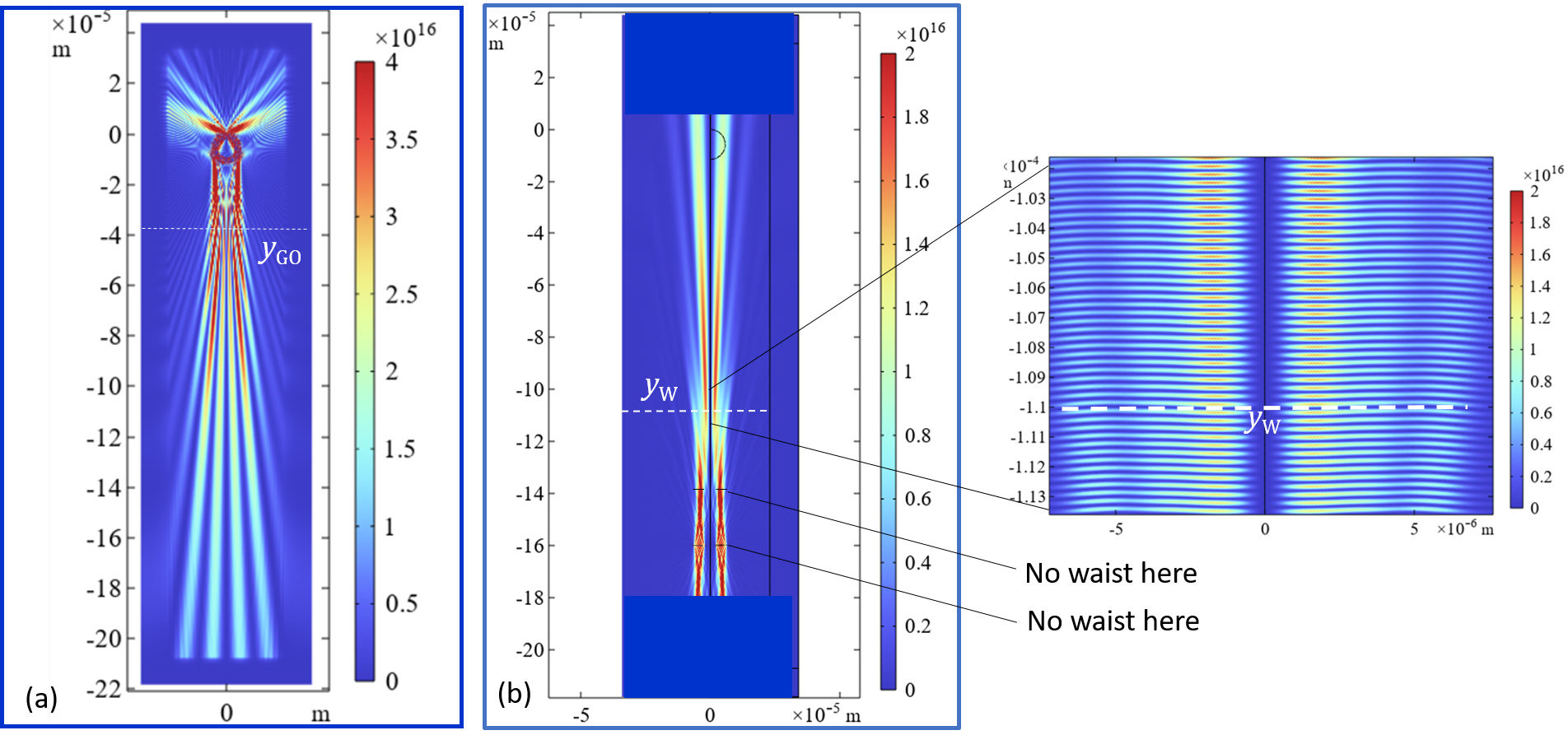}
\caption{(a) Intensity color map for the case $R=10.45\lambda$, $n=1.7$.  
(b) Intensity color map for the backward beam in the domain of the waist.}
\label{Pic3new}
\end{figure*}

The same observations can be done for all size parameters  $R/\lambda$ in the range $5-11$ we have studied with the step $0.05$. Besides of the resonant values of $n$, large $n$ correspond to the novel scenario of imaging and small $n$ -- to the conventional scenario. In all cases the resolution is better in the novel regime. 
For example, for $R/\lambda=10.45$ $n=1.3$ the scenario of imaging is illustrated by Fig. \ref{Pic4new}. In Fig. \ref{Pic4new}(a) we see the conventional scenario of imaging with the phase center of the imaging beam located at $y_{GO}=8\, \mu$m. In Fig. \ref{Pic4new}(b) we see the backward beam waist located at $y_{GO}=8.5\, \mu$m.
The ultimate resolution is equal in this case $\delta_{\rm min}=0.97\lambda$.  Fig. \ref{Pic3new} illustrated the case $R/\lambda=10.45$ $n=1.7$, when we have 
$y_{GO}=-41\, \mu$m, $y_W=-110\, \mu$m. In spite of quite strong magnification ($M\approx 12$) the resolution is modest: $\delta_{\rm min}\approx 0.54\lambda$ due to the rather wide waist that gives for the VS the effective width $d_V=5.4\, \mu$m. However, this resolution is also better than that achieved in the conventional scenario.

Note, that in accordance to our previous studies the qualitative 
applicability of the geometrical optics to the formation of the imaging beam was confirmed only for $R>15\lambda$. For so big microparticles 
our present study is hardly feasible. It demands to simulate the optical path 
longer than $1000\lambda$ in order to properly choose the value of $y_S$ i.e. to retrieve the backward beam reliably.

\section{Conclusions}

In this article, we have reported the result of the comparative theoretical study of two scenarios of nanoimaging granted by a simple dielectric microparticle 
(cylinder) whose refractive index was varied in a broad interval.  
In the first (known) scenario the magnified virtual source is formed in front of the microparticle. This is so when $n\sim 1.3-1.4$.    
in the second (new) scenario it is formed far behind it. This scenario holds when $n\sim 1.5-1.6$. 
We numerically studied the far-field imaging for microparticles with size parameters in the range $R/\lambda$ from $4$ to $11$ and found that in all non-resonant cases the new scenario grants higher magnification and finer resolution, and is, therefore, more promising than the known one.
The method of our study is based on the use of the backward beam propagating in free space. This novel method offers the virtual source -- an effective spread source creating at very large distances (much larger than the Rayleigh range) the same imaging beam as that created by the real source and the microparticle.   
The resolution of the VS and its magnification compared to the real source grants us the opportunity to simulate the superresolution without involving the 
objective of a microscope and considering the optical paths shorter than $1000\lambda$. It allows us to use the full-wave (COMSOL) simulations. 

The drawback of our study is heuristic criterion of resolution of two adjacent VSs that does not take into account the nonzero angle between the optical axes of two imaging beams created by two adjacent sources. We assume these optical axes parallel. This approach gives only a pessimistic estimate for the resolution of virtual sources. However, even these pessimistic estimates can be improved if we take into account the substrate (in \cite{Reza1} it was shown that the impact of the substrate is favorable for superresolution). The replacement of a microcylinder by a microsphere will also improve the ultimate resolution.

Of course, our final goal is the experimental check of our theoretical predictions. Since the experimenters prefer to use the same techniques 
as were used in the cited papers on superresolution granted by a glass microsphere we cannot rely on specific values of $n$ and $R/\lambda$, granting two different scenarios of imaging for microcylinders. We have to carry out 3D simulations and find corresponding values for microspheres.  
If we manage to do it with a supercomputer we will look for an interested team of opticians.  
Experimental results for the novel scenario are not known, and the theory predicts finer resolution namely for this regime. 
Therefore, we think that such the experiment is very important and may open a new chapter in the label-free 
far-field nanoimaging offered by dielectric microspheres.


\section{Disclosures}
The authors declare no conflicts of interest.

\bibliography{paper}

\end{document}